# Hidden magnetism at the pseudogap critical point of a high temperature superconductor

Mehdi Frachet[1]†, Igor Vinograd[1]†, Rui Zhou[1,2], Siham Benhabib[1], Shangfei Wu[1], Hadrien Mayaffre[1], Steffen Krämer[1], Sanath K. Ramakrishna[3], Arneil P. Reyes[3], Jérôme Debray[4], Tohru Kurosawa[5], Naoki Momono[6], Migaku Oda[5], Seiki Komiya[7], Shimpei Ono[7], Masafumi Horio[8], Johan Chang[8], Cyril Proust[1], David LeBoeuf[1]*, Marc-Henri Julien[1]*

[1] Univ. Grenoble Alpes, INSA Toulouse, Univ. Toulouse Paul Sabatier, EMFL, CNRS, LNCMI, 38000 Grenoble, France

[2] Institute of Physics, Chinese Academy of Sciences and Beijing National Laboratory for Condensed Matter Physics, Beijing 100190, China

[3] National High Magnetic Field Laboratory, Florida State University, Tallahassee, FL 32310, USA

[4] Université Grenoble Alpes, CNRS, Grenoble INP, Institut Néel, Grenoble, France

[5] Department of Physics, Hokkaido University, Sapporo 060-0810, Japan

[6] Muroran Institute of Technology, Muroran 050-8585, Japan

[7] Central Research Institute of Electric Power Industry, Yokosuka, 240-0196, Japan

[8] Department of Physics, University of Zurich, CH-8057 Zurich, Switzerland

† These authors contributed equally to this work.

*Correspondence to: david.leboeuf@lncmi.cnrs.fr, marc-henri.julien@lncmi.cnrs.fr.

**The mysterious pseudogap phase of cuprate superconductors ends at a critical hole doping level $p^*$ but the nature of the ground state below $p^*$ is still debated. Here, we show that the genuine nature of the magnetic ground state in $La_{2-x}Sr_xCuO_4$ is hidden by competing effects from superconductivity: applying intense magnetic fields to quench superconductivity, we uncover the presence of glassy antiferromagnetic order up to the pseudogap boundary $p^* \approx 0.19$, and not above. There is thus a quantum phase transition at $p^*$, which is likely to underlie high-field observations of a fundamental change in electronic properties across $p^*$. Furthermore, the continuous presence of quasi-static moments from the insulator up to $p^*$ suggests that the physics of the doped Mott insulator is relevant through the entire pseudogap regime and might be more fundamentally driving the transition at $p^*$ than just spin or charge ordering.**

Extensive studies of the cuprates (1) have shown that, after three-dimensional Néel order disappears upon hole doping ($p$), there are still remains of spin order at low temperature ($T$), in the form of a glass-like freezing of incommensurate antiferromagnetic correlations (2-6). In $La_{2-x}Sr_xCuO_4$, this "antiferromagnetic glass" is favored by charge-stripe ordering around $p = x = 0.12$ and is clearly observed up to a doping $p_{sg} \approx 0.135$ (Fig. 1a) (2,3).

Persistence of slow spin fluctuations and impurity-induced freezing up to $p^* \approx 0.19$ (3) has led to hypothesize that ground state of the pseudogap regime is an antiferromagnetic glass (3) and that antiferromagnetic correlations exist only below $p^*$, that is, within the pseudogap phase (7). The results are, however, controversial (4) and interpretation may be ambiguous as impurity doping intrinsically favors randomness and freezing while at the same time weakening superconductivity. Furthermore, a scenario connecting glassy freezing with the pseudogap faces difficulties, among which are the persistence of antiferromagnetic correlations above $p^*$ (8) and the disappearance of spin freezing well below $p^*$ in most cuprates ($p_{sg}$ = 0.08 in YBa$_2$Cu$_3$O$_y$, refs 5,6).

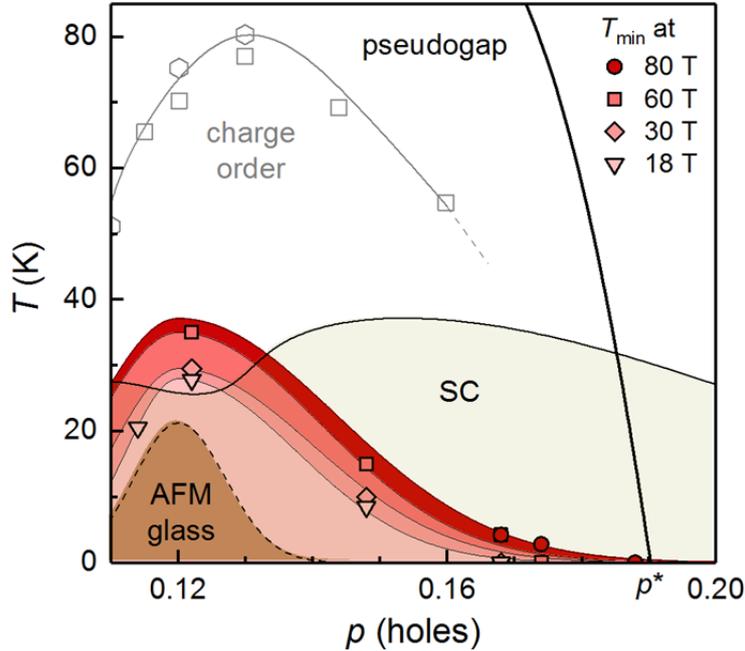

**Fig. 1. Quasi-static magnetism in the pseudogap state of La$_{2-x}$Sr$_x$CuO$_4$.** Temperature – doping phase diagram representing $T_{min}$, the temperature of the minimum in the sound velocity, at different fields. Since superconductivity precludes the observation of $T_{min}$ in zero-field, the dashed line (brown area) represents the extrapolated $T_{min}(B=0)$. While not exactly equal to the freezing temperature $T_f$ (see Fig. 2), $T_{min}$ is closely tied to $T_f$ and so is expected to have the same doping dependence, including a peak around $p$ = 0.12 in zero/low fields (ref. 2). Onset temperatures of charge order are from ref. 33 (squares) and 35 (hexagons).

Here, we follow a different approach, without impurity doping, that sheds new light on these fundamental issues: quenching superconductivity with high magnetic fields reveals that the antiferromagnetic glass of La$_{2-x}$Sr$_x$CuO$_4$ extends from the weakly doped insulator up to the pseudogap boundary $p^*$, when not hindered by superconductivity. Specifically, previous neutron scattering studies showed that a field $B$ enhances static magnetism for $p \approx 0.10 – 0.12$ and even induces it for $p \approx 0.145$ (9-11) but not at higher doping where only the finite-energy spectrum is affected (12,13). Here, using much higher fields, we discover that static or quasi-static magnetism actually persists well above $p \approx 0.145$ but not across the whole phase diagram: in fact, only up to the critical doping $p^* \approx 0.19$ of the pseudogap phase.

In order to provide a benchmark for measurements near $p^*$, we first report $^{139}$La nuclear magnetic resonance (NMR) and ultrasound results for the doping $p$ = 0.148 at which magnetism should be field-dependent (13-15) (see Methods for experimental and sample details). Glassy freezing is typically seen in NMR as a broad peak in the nuclear spin-lattice relaxation rate $1/T_1$ vs. $T$ when the inverse correlation time of spin fluctuations $\tau_c^{-1}$ matches the NMR frequency $\omega_{NMR}$ (6,14-17). Such a peak, defining a freezing temperature $T_f$ at the NMR timescale, is seen very clearly in our high-field data and it disappears at low fields (Fig. 2). The spatial heterogeneity that typifies spin-glasses and $La_{2-x}Sr_xCuO_4$ with $x < 0.135$ (14-17) is also present here, as shown by the large distribution of $T_1$ values (see Materials and Methods and Fig. S8).The high-field data for the sound velocity $\Delta v/v$ of the transverse mode $(c_{11} - c_{12})/2$ (Fig. 2) is equally unambiguous for $p = 0.148$: a softening (decrease) in the sound velocity $\Delta v/v$, followed by a hardening (increase), is observed upon cooling, exactly as in canonical spin-glasses (18). This reflects the temperature dependence of the dynamical magnetic susceptibility which also depends on $\tau_c$. The resulting minimum in $\Delta v/v$ occurs at a temperature $T_{min}$ slightly higher than $T_f$ defined from NMR. Notice that what we detect here is not a true phase transition but an apparent freezing at the experimental timescale (in the MHz range for both NMR and ultrasound): the moments continue to slow down (typically exponentially) on cooling below $T_f$ and become truly static only at much lower temperature.

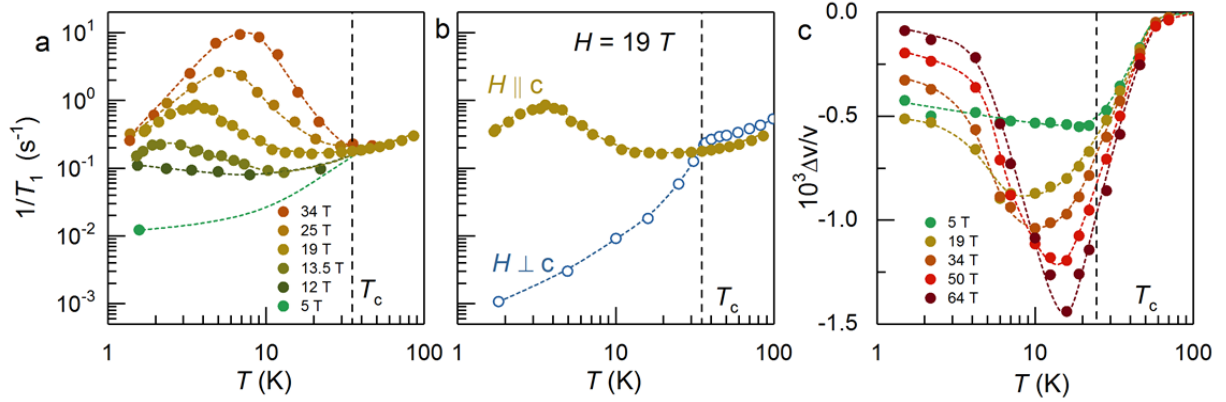

**Fig. 2. NMR and ultrasound signatures of field-induced glassy freezing in $La_{1.852}Sr_{0.148}CuO_4$. a**, $^{139}La$ $1/T_1$ *vs.* temperature at different fields (applied parallel to the *c* axis). The broad peak in $1/T_1$ is characteristic of continuous slowing down of spin fluctuations, with a correlation time diverging towards $T \approx 0$ (typically $\tau_c = \tau_\infty \exp(E_0/k_B T)$ with an activation energy $E_0$). The peak amplitude decreases and its width increases upon decreasing *B* such that the peak is no longer discerned at low fields. **b**, $^{139}La$ $1/T_1$ vs. temperature at $B = 19$ T applied parallel or perpendicular to the *c* axis. For $B \perp c$, $1/T_1$ drops strongly below $T_c$, as expected for a superconductor, without any sign of enhanced spin fluctuations. **c**, Sound velocity $\Delta v/v$ in the $(c_{11} - c_{12})/2$ mode *vs.* temperature at different fields (applied parallel to the *c* axis), after subtraction of a lattice background (see Methods). In spin-glasses, the softening (decrease of $\Delta v/v$ upon cooling) mirrors the increase of the magnetic susceptibility above the freezing temperature $T_f$ (18). There is a minimum in $\Delta v/v$ at $T_{min}$ before the lattice hardens upon further cooling through $T_f$. Such a dip in $\Delta v/v$ has been reported earlier in $La_{1.86}Sr_{0.14}CuO_4$ at lower fields (19) and was interpreted as the result of a competition between a structural instability (producing a softening above $T_c$) and superconductivity (producing a hardening below $T_c$). Our measurements up to 64 T show that the minimum in $\Delta v/v$ cannot be due to superconductivity as it gets deeper and is shifted to higher temperature as the field increases. Dashed lines are guides to the eye.

We now investigate whether glassy freezing can be detected at higher doping. It turns out that, for doping levels from $p = 0.155$ to $0.188$, data from both techniques are qualitatively similar to that for $p = 0.148$ (Fig. 3). Upon cooling at low temperatures in high fields (Fig. 3), a softening is observed in ultrasound, though at temperatures lower than for $p = 0.148$. Also, even though a peak in $1/T_1$ vs. $T$ is not observed for $p = 0.171$ at our highest field, $1/T_1$ values are much larger than expected from an extrapolation of the normal state values, showing that the effect of the field is not just to close the superconducting gap (in Fig. 3, we plot $1/T_1 T$ instead of $1/T_1$ in order to better highlight the difference with the normal state where $1/T_1 T$ is constant just above $T_c$).

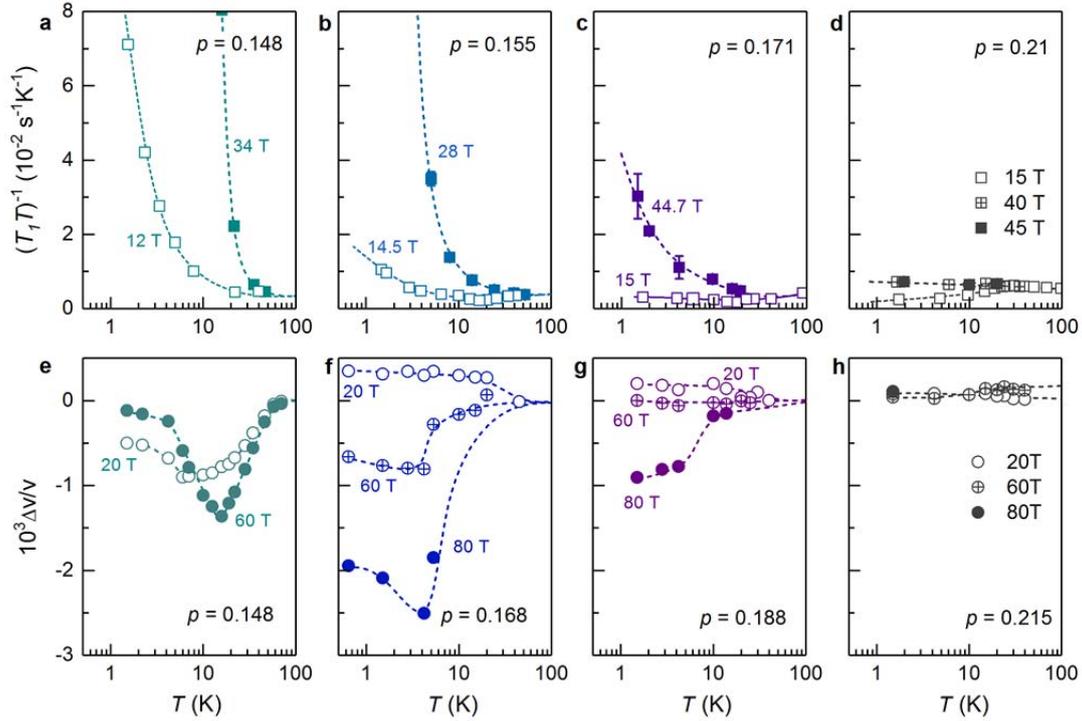

**Fig. 3. Doping dependence of spin freezing in high fields. a-d**, $^{139}$La $1/T_1T$ vs. temperature at different fields and hole doping levels. For $p = 0.21$, $1/T_1$ saturates in high fields at values extrapolated from $T > T_c$, meaning that all of the field effect in $1/T_1$ arises from the closure of the superconducting gap. Notice that $1/T_1$ values for this compound are extrinsically high due to lattice fluctuations at low temperature (Fig. S4). **e-h**, sound velocity $\Delta v/v$ vs. temperature at different fields and hole doping levels. For $p = 0.168$ and $0.188$, $\Delta v/v$ at $B = 20$ T increases upon cooling and saturates at low temperature. This behaviour is explained by the coupling of superconductivity with the lattice (see Methods). With increasing $B$, this superconducting contribution to the sound velocity is reduced and a lattice softening develops. For $p = 0.215$, $\Delta v/v$ shows almost no temperature dependence up to 80 T and down to 1.5 K. Notice that, while we detect slow spin fluctuations up to $p = 0.188$, freezing at the NMR or ultrasound timescales (both in the MHz range) is not reached in two of our datasets: for $p = 0.171$, $1/T_1T$ is anomalously enhanced in high fields (panel 3c and Fig. 4a) but there is no peak of $1/T_1$ vs. $T$ even at 45 T. For $p = 0.188$, we observe a lattice softening but not the hardening that signals the frozen state, at least down to 1.5 K in a field of 80 T. Dashed lines are guides to the eye.

At base temperature (~1.5 K), both $1/T_1$ and $(\Delta v/v)^{-1}$ grow with field (Fig. 4a,b) but the field scale required to observe this increase grows with doping (see also Fig. S9). This is visualised in the doping dependence of the field scale $B_{slow}$ that characterises the onset of slow spin fluctuations (Fig. 4c). Typically, $B_{slow}$ is the field above which an elastic response should appear in neutron scattering. At $p \approx 0.17$, $B_{slow} \approx 30$ T is already as high as ~2/3 of the upper critical field $B_{c2}$.

The above results are remarkably consistent with theories (20) in which spin order competes fiercely with superconductivity: the competing order is enhanced in and around vortex cores and progressively takes over superconductivity as the field, and thus the vortex density, increases. The striking absence of spin-freezing when $B$ is aligned within the CuO$_2$ planes (Fig. 2b and Fig. S9) ascertains that magnetism is induced by the weakening of superconductivity, not by the field itself. Also, the existence of a doping-dependent field scale $B_{slow}$ (most clearly seen in ultrasound data – Fig. 4b) is suggestive of a quantum critical point shifted towards lower doping levels by superconductivity (*20*).

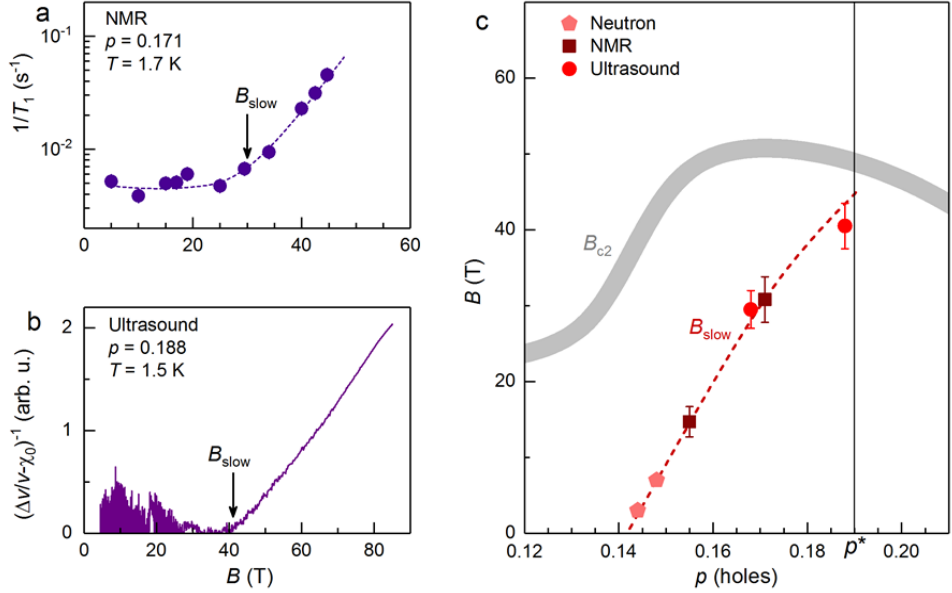

**Fig. 4. Field dependence of quasi-static magnetism. a**, Field dependence of $^{139}$La $1/T_1$ at $T = 1.7$ K for $p = 0.171$ (see Fig. S8 for additional doping levels). The arrow points the onset field of slow spin fluctuations $B_{slow}$ (defined so as to match the neutron onset field for $p = 0.148$ (*11*), as explained in the Methods section). **b**, Field dependence of $(\Delta v/v - \chi_0)^{-1}$ for $p = 0.188$ and $T = 1.5$ K ($\geq T_{min}$), where $\chi_0$ is a doping-dependent constant (see Fig. S8 for additional doping levels). $\Delta v/v$ is almost field independent at low fields but above a doping-dependent onset field $B_{slow}$ (pointed by an arrow), it shows a clear $1/B$ dependence. Note that, within error bars, $B_{slow}$ has no temperature dependence between 0.6 and 4.2 K, for $p = 0.168$ (Fig. S3). **c**, Doping dependence of $B_{slow}$ and $B_{c2}$ (see Fig. S10). Dashed lines are guides to the eye.

There is, however, a fundamental aspect of the data that has not been anticipated by available theories: for $p \approx 0.21 > p^*$, the results (Figs. 3 and Fig. S9) are qualitatively different, with a modest field dependence that is entirely understood from just the closure of the superconducting gap. This is our main finding, summarized in a phase diagram (Fig. 1): the field-dependent spin freezing ends at $p^*$, which means that, once superconductivity is quenched in high fields, the pseudogap boundary at zero temperature corresponds to a quantum phase transition from glassy antiferromagnetic order to a correlated metal with only short-lived antiferromagnetism.

We now explore different implications of this discovery.

In recent years, remarkable electronic changes associated with the pseudogap boundary have been revealed in high fields (*23-26*). These have suggested that $p^*$ is a quantum critical point and, by extension, that the pseudogap is characterized by (symmetry-breaking or topological) order (*1,21,22* and references therein). However, the here-revealed quantum phase transition is likely to underlie the observed changes at $p^*$, thus showing that quantum critical behaviour may arise from a low-temperature order, which does not require the pseudogap state itself to be an ordered state. Specifically, our results unambiguously show that the effect of the field can no longer be ignored when interpreting high-field measurements in the vicinity of $p^*$ (*23-27*): in fields comparable to $B_{c2}$, the antiferromagnetic moments should fluctuate slowly enough to strongly enhance scattering of low-energy electronic states, thus impacting transport properties, and they might be correlated over long-enough distances to even reconstruct the Fermi-surface (*e.g.* the correlation length is $\xi_{AF} > 100$ lattice spacings in La$_{1.88}$Sr$_{0.12}$CuO$_4$, ref. 9). We also conjecture that the negative thermal Hall coefficient (28) is related to properties of these slow antiferromagnetic moments observed in a similar range of temperatures and doping levels. Furthermore, the recently-reported quantum critical behaviour in La$_{1.6-x-y}$Nd$_{0.4}$Sr$_x$CuO$_4$ (26) likely coincides with a similar transition at $T = 0$ since this compound shows analogous field-dependent spin and charge stripe correlations as La$_{2-x}$Sr$_x$CuO$_4$ (29).

In addition to potentially explaining an important part of the pseudogap phenomenology, the results fundamentally reveal that, if not masked by superconductivity, the same local-moment antiferromagnetism as found on the verge of the Néel phase survives throughout the pseudogap state. This observation clearly favours theories in which the pseudogap state and its various associated electronic orders originate from strong correlation physics fundamentally rooted in the doped Mott insulator.

An antiferromagnetic quantum critical point hidden by superconductivity has previously been observed in a heavy-fermion metal (30) and in an electron-doped cuprate (31). The situation would be similar here if spin order appeared sharply at the pseudogap onset temperature $T^*$. However, this is not the case: spin fluctuations progressively slow down on cooling and freeze at temperatures that are one to two orders of magnitude lower than $T^*$. Furthermore, the antiferromagnetic glass of La$_{2-x}$Sr$_x$CuO$_4$ is, to a large part, viewed as a consequence of charge-stripe (or of pair-density-wave) order for $p \approx 0.12$ (Fig. 1 and ref. 21). Then, a natural suspicion is that the former disappears because the latter vanishes at $p^*$. Within this perspective (32) and neglecting possible complications such as phase separation, the quantum phase transition at $p^*$ would be a transition from a stripe-ordered metal to a correlated but uniform metal. This is not inconsistent with recent data (33) but since the antiferromagnetic glass may in principle exist without charge order (it does apparently so at low doping), experiments testing the disappearance of charge order near $p^*$ in La$_{2-x}$Sr$_x$CuO$_4$ are urged for.

Regardless of the exact role of spin and charge orders in the transition at $p^*$, the continuous presence of quasi-static moments from the insulator at $p \approx 0.03$ up $p^* \approx 0.19$ suggests that these moments are inherited from the Mott insulator (34) and, as such, are localized. We note that the carrier density $p$ found below $p^*$ (24,25) is consistent with moments localized at Cu sites. If those local moments were still present above $p^*$, they would ultimately slow down at low temperature and be detected, unless they enter a spin-singlet ground-state above $p^*$ which seems unlikely. Instead, the moments becoming mostly itinerant above $p^*$, as inferred from the increase in carrier density from $p$ to $1+p$ (24,25), could be detrimental to glassy freezing. The pseudogap phase would then be distinct from the correlated metal found for $p > p^*$ by a unique ability to sustain such moments. In that sense, the transition at $p^*$ could be primarily associated with the loss of Mott physics, of which the simultaneous disappearance of spin and/or charge order would be consequences.

**References**

1. B. Keimer, S. A. Kivelson, M. R. Norman, S. Uchida and J. Zaanen, Nature **518**, 179–186 (2015). *From quantum matter to high-temperature superconductivity in copper oxides*.

2. M.-H. Julien, Magnetic order and superconductivity in La$_{2-x}$Sr$_x$CuO$_4$: a review. Physica B **329**, 693 (2003).

3. C. Panagopoulos *et al.* Solid State Communications **126**, 47–55 (2003). *Low-frequency spins and the ground state in high-$T_c$ cuprates*.

4. Risdiana *et al.* Phys. Rev. B. **77**, 054516 (2008). *Cu spin dynamics in the overdoped*

regime of $La_{2-x}Sr_xCu_{1-y}Zn_yO_4$ probed by muon spin rotation.

5. D. Haug *et al.* New. J. Phys. **12**, 105006 (2010). *Neutron scattering study of the magnetic phase diagram of underdoped $YBa_2Cu_3O_{6+x}$.*

6. T. Wu *et al.* Phys. Rev. B **88**, 014511 (2013). *Magnetic-field-enhanced spin freezing on the verge of charge ordering in $YBa_2Cu_3O_{6.45}$.*

7. J. L. Tallon, J. W. Loram, and C. Panagopoulos, J. Low Temp. Phys. **131**, 387–394 (2003). *Pseudogap and quantum-transition phenomenology in HTS cuprates.*

8. Y. Li *et al.*, Phys. Rev. B **98**, 224508 (2018). *Low-energy antiferromagnetic spin fluctuations limit the coherent superconducting gap in cuprates.*

9. B. Lake *et al.* Nature **415**, 299–302 (2002). *Antiferromagnetic order induced by an applied magnetic field in a high-temperature superconductor.*

10. B. Khaykovich *et al.* Phys. Rev. B **71**, 220508R (2005). *Field-induced transition between magnetically disordered and ordered phases in underdoped $La_{2-x}Sr_xCuO_4$.*

11. J. Chang *et al.* Phys. Rev. B **78**, 104525 (2008). *Tuning competing orders in $La_{2-x}Sr_xCuO_4$ cuprate superconductors by the application of an external magnetic field.*

12. B. Lake *et al.* Science **291**, 1759–1762 (2001). *Spins in the vortices of a high-temperature superconductor.*

13. J. M. Tranquada *et al.* Phys. Rev. B **69**, 174507 (2004). *Evidence for an incommensurate magnetic resonance in $La_{2-x}Sr_xCuO_4$.*

14. V. F. Mitrović *et al.* Phys. Rev. B **78**, 014504 (2008). *Similar glassy features in the $^{139}La$ NMR response of pure and disordered $La_{1.88}Sr_{0.12}CuO_4$.*

15. A. Arsenault *et al.* Phys. Rev. B **97**, 064511 (2018). *$^{139}La$ NMR investigation of the charge and spin order in a $La_{1.885}Sr_{0.115}CuO_4$ single crystal.*

16. S.-H. Baek, A. Erb and B. Büchner, Phys. Rev. B **96**, 094519 (2017). *Low-energy spin dynamics and critical hole concentrations in $La_{2-x}Sr_xCuO_4$ ($0.07 \leq x \leq 0.2$) revealed by $^{139}La$ and $^{63}Cu$ nuclear magnetic resonance.*

17. N. J. Curro, et al. Phys. Rev. Lett. **85**, 642 (2000). *Inhomogeneous Low Frequency Spin Dynamics in $La_{1.65}Eu_{0.2}Sr_{0.15}CuO_4$.*

18. P. Doussineau, A. Levelut, M. Matecki, M. Renard, J. P. Schön, W. EPL **3**, 251 (1987). *Acoustic and magnetic studies of an insulating spin glass.*

19. M. Nohara et al. Phys. Rev. Lett. **70**, 3447 (1993). *Interplay between lattice softening and high-$T_c$ superconductivity in $La_{1.86}Sr_{0.14}CuO_4$.*

20. E. Demler, S. Sachdev and Y. Zhang, Phys. Rev. Lett. **87**, 067202 (2001). *Spin-ordering quantum transitions of superconductors in a magnetic field.*

21. E. Fradkin, S. A. Kivelson, J. M. Tranquada, Rev. Mod. Phys. **87**, 457–482 (2015). *Theory of intertwined orders in high temperature superconductors.*

22. M. S. Scheurer, S. Chatterjee, W. Wu, M. Ferrero, A. Georges, and S. Sachdev, PNAS **115**, E3665–E3672 (2018). *Topological order in the pseudogap metal.*

23. B. J. Ramshaw *et al.* Science **348**, 317–320 (2015). *Quasiparticle mass enhancement approaching optimal doping in a high-$T_c$ superconductor.*

24. S. Badoux *et al.*, Nature **531**, 210–214 (2016). *Change of carrier density at the pseudogap critical point of a cuprate superconductor.*

25. F. Laliberté *et al.* http://arxiv.org/abs/1606.04491. *Origin of the metal-to-insulator crossover in cuprate superconductors.*

26. B. Michon *et al.* Nature **567**, 218–222 (2019). *Thermodynamic signatures of quantum criticality in cuprate superconductors.*

27. R. A. Cooper et al. Science **323**, 603–607 (2009). *Anomalous criticality in the electrical resistivity of $La_{2-x}Sr_xCuO_4$.*

28. G. Grissonnanche et al. Nature **571**, 376–380 (2019). *Giant thermal Hall conductivity from neutral excitations in the pseudogap phase of cuprates.*

29. M. Hücker et al. Phys. Rev. B **87**, 014501 (2013). *Enhanced charge stripe order of superconducting $La_{2-x}Ba_xCuO_4$ in a magnetic field.*

30. T. Park *et al.* Nature **440**, 65–68 (2006). *Hidden magnetism and quantum criticality in the heavy fermion superconductor $CeRhIn_5$.*

31. H. J. Kang *et al.* Nature **423**, 522–525 (2003). *Antiferromagnetic order as the competing ground state in electron-doped $Nd_{1.85}Ce_{0.15}CuO_4$.*

32. K. Fujita *et al.* Science **344**, 612–616 (2014). *Simultaneous transitions in cuprate momentum-space topology and electronic symmetry breaking.*

33. J. J. Wen et al. Nat. Commun. **10**, 3269 (2019). *Observation of two types of charge density wave orders in superconducting $La_{2-x}Sr_xCuO_4$.*

34. O. Parcollet and A. Georges, Phys. Rev. B **59**, 5341–5360 (1999). *Non-Fermi-liquid regime of a doped Mott insulator.*

35. T. P. Croft, C. Lester, M. S. Senn, A. Bombardi, and S. M. Hayden. Phys. Rev. B **89**, 224513 (2014). *Charge density wave fluctuations in $La_{2-x}Sr_xCuO_4$ and their competition with superconductivity.*

**Acknowledgments:**

Part of this work was performed at the LNCMI, a member of the European Magnetic Field Laboratory. A portion of this work was performed at the National High Magnetic Field Laboratory, which is supported by the National Science Foundation Cooperative Agreement No. DMR-1644779 and the State of Florida. Work in Grenoble was supported by the Laboratoire d'Excellence LANEF (ANR-10-LABX-51-01) and contract ANR-14-CE05-0007. Work in Toulouse was supported through the EUR grant NanoX n° ANR-17-EURE-0009. Work in Beijing was supported by the National Natural Science Foundation of China (grants No. 11674377, No. 11634015, No. 11974405) and MOST grants (No. 2016YFA0300502 and No. 2017YFA0302904). Work in Zürich was supported by the Swiss National Science Foundation. S. O. acknowledges support from JSPS KAKENHI Grant Number JP17H01052.

**Author contributions:** M.F., S.B., S.W., C.P. and D.L. performed the ultrasound experiments. I.V., R.Z., H.M., S.K., S.K.R., A.R. and. M.-H.J. performed the NMR experiments. M.F. and I.V. analyzed experimental data with suggestions from D.L. and M.-H.J.. T.K., N.M., M.O., S.K., S.O., M.H. and J.C provided single crystals. M.F. and J.D. cut precisely-oriented single crystals. D.L. and M.-H.J. supervised the project and wrote the manuscript with suggestions from all authors.

## Supplementary Material

### Samples

High quality $La_{2-x}Sr_xCuO_4$ (LSCO) single crystals were grown by the traveling solvent floating zone method. NMR samples with $x = p = 0.171$ and $x = p = 0.210$ were cut from the same rods as the US samples with $p = 0.168$ and $p = 0.215$, respectively (see below for the estimation of $p$ values).

### Determination of doping level

The hole doping $p$, which is considered to be equal to the Sr concentration $x$ in the absence of oxygen off-stoichiometry, has been determined by measuring $T_{st}$, the temperature of the structural transition from the high-$T$ tetragonal (HTT) phase to the low-$T$ orthorhombic (LTO) phase (36,37), either by NMR or sound velocity (Fig. S2). $T_{st}$ provides a more accurate measurement of doping than $T_c$ because it varies more strongly with hole doping and is less sensitive to defects. In NMR samples, $T_{st}$ was measured using the spin-lattice relaxation rate $1/T_1$ of $^{139}La$, which shows a peak at the transition (Fig. S5 and refs. 15,16). In US samples it was detected by measuring the $(c_{11} - c_{12})/2$ mode which shows an anomaly at the transition (Fig. S1).

Since the HTT to LTO transition line goes to 0 at $p \approx 0.21$, the doping of samples with Sr content above this value is assessed using the superconducting critical temperature $T_c$. $T_c$ of NMR samples was measured by tracking the resonance frequency of the NMR tank circuit, which showed a strong and very sharp change at $T_c$ for all samples. In US samples, an anomaly at $T_c$ is detected in the temperature dependence of the sound velocity $\Delta v/v$ (see Fig. S3).

Table S1 summarizes the properties of the single crystals used in this study. Typical (relative) uncertainty on doping is ±0.002 hole/Cu, except for $p \geq 0.21$ where it is ±0.005 hole/Cu.

### Ultrasound measurements

Transverse ultrasonic waves at typical frequencies ranging from 100 MHz to 200 MHz were generated using commercial $LiNbO_3$ 41° X-cut transducers glued on oriented, polished, and cleaned surfaces. We used a special set up that allows to orient the crystal in a Laue diffractometer and to transfer the crystal on a wire saw while conserving the orientation within a typical precision of one degree.

A standard pulse-echo technique with phase comparison was used to measure sound velocity variation $\Delta v/v$ (38). The experiments were performed at the LNCMI Toulouse in pulsed fields up to 86 T. A high-speed acquisition system was used to record the evolution of the phase of the acoustic echoes during the magnetic field pulse. When possible, good reproducibility was checked at different frequencies and on different echoes. Data on the upsweep and downsweep of the magnetic field pulses showed good overlap, indicating constant temperature of the sample during the pulse. The field dependence of the sound velocity at different temperatures is shown for all samples in Fig. S4.

### Contributions to the sound velocity

The measurements in pulsed fields give access to the field-induced sound velocity, $\Delta v/v(B)$, at different temperatures (Fig. S4). By performing constant-field cuts in these field sweeps, we obtain the temperature dependence of the field-dependent sound velocity. In order to obtain the complete temperature dependence of the sound velocity

$\Delta v/v$ at different fields (Fig 2 and 3), we need to add the zero-field electronic sound velocity, $\Delta v/v(B=0)_e$ :

$$\Delta v/v = \Delta v/v(B) + \Delta v/v(B=0)_e \qquad \text{Eq. E1}$$

$\Delta v/v(B=0)_e$ includes the influence of superconductivity and magnetism on the lattice. $\Delta v/v(B=0)_e$ is extracted from the raw data of the zero field temperature dependent sound velocity, $\Delta v/v(B=0)$ which also contains a background component, $\Delta v/v(B=0)_{background}$. This background originates from the natural hardening of the lattice upon cooling of the sample. This background is fitted, within a temperature range where neither magnetism nor superconductivity make significant contribution to $\Delta v/v(B=0)$, with an empirical formula:

$$\Delta v/v(B=0)_{background} = c - \frac{s}{\exp\left(\frac{t}{T}\right)-1} \qquad \text{Eq. E2}$$

Eq. E2 has been shown to describe accurately the lattice contribution to the elastic constant in a wide variety of systems (39), including cuprates (40). Fig. S3 shows the zero field sound velocity data $\Delta v/v(B=0)$, the background fit $\Delta v/v(B=0)_{background}$, and the extracted zero field electronic contribution $\Delta v/v(B=0)_e$.

## NMR measurements

Experiments were performed using standard spin-echo techniques and home-built heterodyne spectrometers in DC fields provided by superconducting and resistive magnets at LNCMI Grenoble and the hybrid magnet at NHMFL, Tallahassee.

The spin lattice relaxation time $T_1$ was determined by fitting the saturation-recovery curve of the $^{139}$La magnetization to a stretched multi-exponential given by eq. E3, which is valid for purely magnetic relaxation of the central line of a nuclear spin 7/2. The stretching exponent $\beta$ phenomenologically accounts for the distribution of $T_1$ values (ref. 14 and references therein) that develops at low temperatures and makes $\beta$ deviate from 1 (Fig. S8). Then, $T_1$ corresponds to the median relaxation rate.

$$\frac{^{139}M_0 - ^{139}M_z(t)}{^{139}M_0} = \frac{1225}{1716}\exp(-\frac{28t}{T_1})^\beta + \frac{75}{364}\exp(-\frac{15t}{T_1})^\beta + \frac{3}{44}\exp(-\frac{6t}{T_1})^\beta + \frac{1}{84}\exp(-\frac{t}{T_1})^\beta$$
Eq. E3

The same expression was used for all temperatures even though quadrupole relaxation (produced by electric-field gradient fluctuations) is present around the structural transition temperature, which causes the exponent $\beta$ to deviate from 1.

## NMR determination of the field scale $B_{slow}$

The only sharp feature in the NMR data is the temperature $T_f$ of the peak in $1/T_1$ vs. $T$ (Fig. 2a). However, it is impossible to define a field above which this peak is present and also impossible to determine whether such a field actually exists. This is because the peak intensity decreases and its width increases upon decreasing the field, therefore making the peak gradually less defined at low fields (Fig. 2a). We also cannot use an arbitrary criterion for the freezing because this would require extrapolation of the data of the $p$ = 0.171 sample, for which the regime where there is a peak in $1/T_1$ vs. $T$ is not reached in the explored range of fields (up to 45 T) and temperatures (down to 1.7 K). Given the impossibility to define the field above which spins are frozen at NMR timescale, we thus resort to the determination of an onset field of slow fluctuations. Given the lack of sharp feature in the field dependence of $1/T_1$ (the upturn at low fields is due to the $1/B$ dependence of $1/T_1$ which arises from $1/T_1 \propto 1/\omega_{NMR}$ and $\omega_{NMR} \propto B$), we define $B_{slow}$ as the field at which the value of $1/T_1$ of a given sample is equal to the value of $1/T_1$ in the $p$ = 0.148 sample at $B$ = 7 T, which is threshold field at which neutron scattering experiments detect a quasi-elastic response (11). So, by construction, $B_{slow}$ = 7 T for $p$ = 0.148. Since the criterion is chosen so that NMR and neutron onset fields match at $p$ = 0.148, there is no NMR point for this sample in Fig. 4c. It turns out that the so-defined $B_{slow}$ in NMR matches the $B_{slow}$ defined from ultrasound (Fig. 4c).

For this determination, we do not use directly the "bare" experimental $T_1$ values because these are biased by two effects: 1) part of the field dependence is due to the reduction of the superconducting gap. 2) As noted above, there is an intrinsic $1/B$ dependence of $1/T_1$ that needs to be corrected. Thus, we determine $B_{slow}$ from $B/T_1$ data from which an approximate field-dependent superconducting component has been subtracted (Figs. S6 and S7). We emphasize that these corrections are small and that they do not affect the conclusions of this paper.

**References**

36. S. Wakimoto *et al.,* J. Phys. Soc. Jpn. **75,** 074714 (2006).
37. H. Takagi *et al.,* Phys. Rev. Lett. **68,** 3777 (1992).
38. B. Lüthi, Physical Acoustics in the Solid State, Springer series in solid state (2004).
39. Y.P. Varshni, Phys. Rev. B **2,** 3952 (1970).
40. M. Nohara *et al.*, Phys. Rev. B **52,** 570 (1995).
41. Y. Ando *et al.,* Phys. Rev. Lett. **75,** 4662 (1995).
42. X. Shi, PhD thesis, Florida State University (2012).
43. P. M. C. Rourke *et al.,* Nat. Phys. **7,** 455-458 (2011).
44. V. Noiseux, Sheerbrooke University (2016). https://www.usherbrooke.ca/sciences/fileadmin/sites/sciences/Physique/Initiation_a_la_recherche/VivianeNoiseux-PIR-Rapport-Article.pdf
45. B. J. Ramshaw *et al.,* Phys. Rev. B **86,** 174501 (2012).
46. G. Blatter *et al.,* Rev. Mod. Phys. **66,** 1125 (1994).
47. H. Won *et al.,* Phys. Rev. B **49,** 1397 (1994).
48. G.-q. Zheng *et al., Phys. Rev. Lett.* **88,** 077003 (2002).
49. B. Keimer *et al.,* Phys. Rev. B **46,** 14034 (1992).
50. M. Reehuis *et al.,* Phys. Rev. B **73,** 144513 (2006).
51. S. Wakimoto *et al.,* J. Phys. Soc. Jpn. **73,** 3413 (2006).

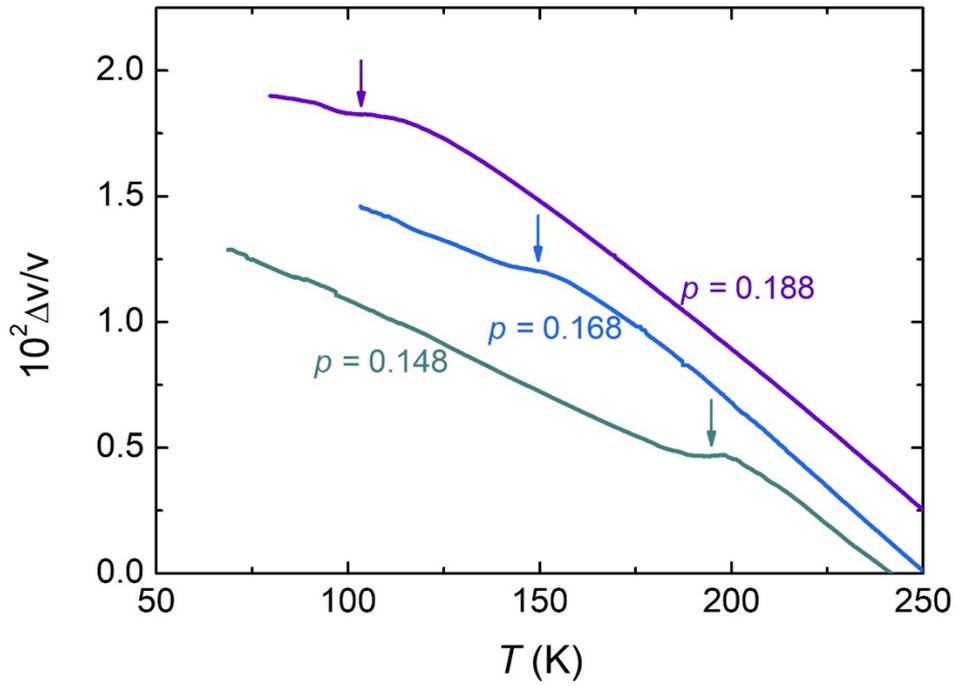

**Fig. S1. Structural transition HTT-LTO seen by sound velocity.**
Sound velocity of the mode $(c_{11}-c_{12})/2$ shows a plateau at $T_{st}$, signaling the high temperature tetragonal (HTT) to low temperature orthorhombic (LTO) phase transition, in agreement with previous report (40). Arrows indicate $T_{st}$ for each doping. For the sample $p = 0.215$ no sign of the structural phase transition is seen down to $T = 7$ K and we therefore used $T_c$ to determine the hole doping of this sample.

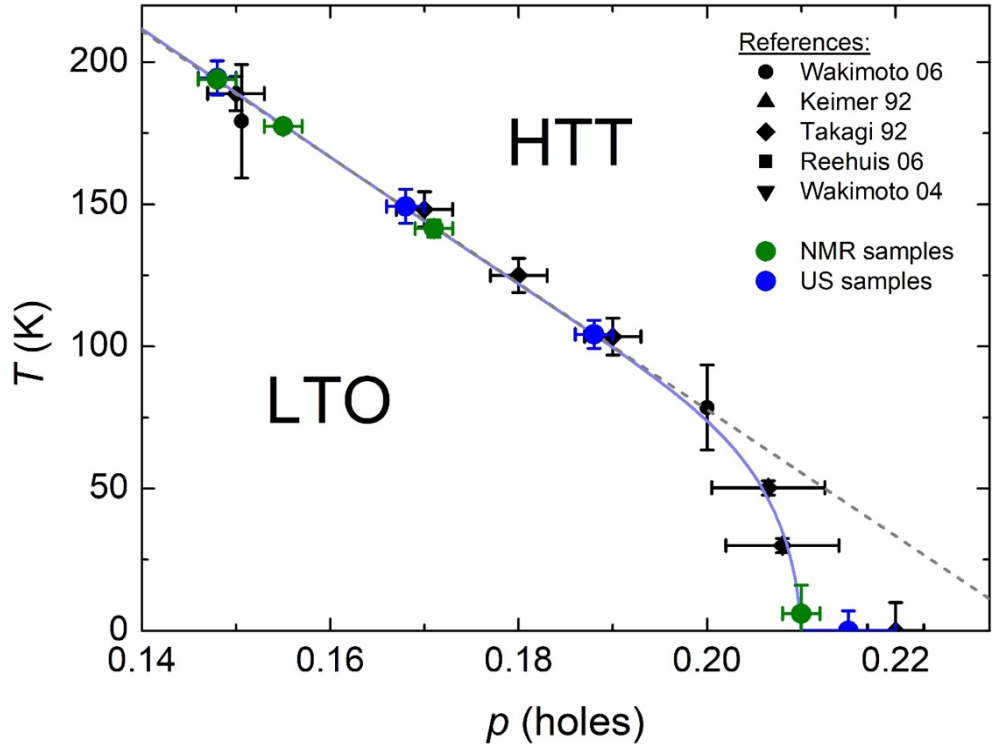

**Fig. S2. Studied samples and their structural transition.**
High temperature tetragonal (HTT) to low temperature orthorhombic (LTO) transition temperature $T_{st}$ of NMR (green) and US (blue) samples plotted as a function of doping. The doping of the NMR and US samples is evaluated by comparing $T_{st}$ with original published data shown as black symbols (36, 37, 49-51), as indicated in the legend. For $p \leq 0.188$, we used the simple linear relation $T_s(p) = 522 - 2221 \times p$ (dashed line) to extract the doping of our samples. The continuous line is a guide to the eye that indicates a critical doping of $p \approx 0.21$ for the structural transition.

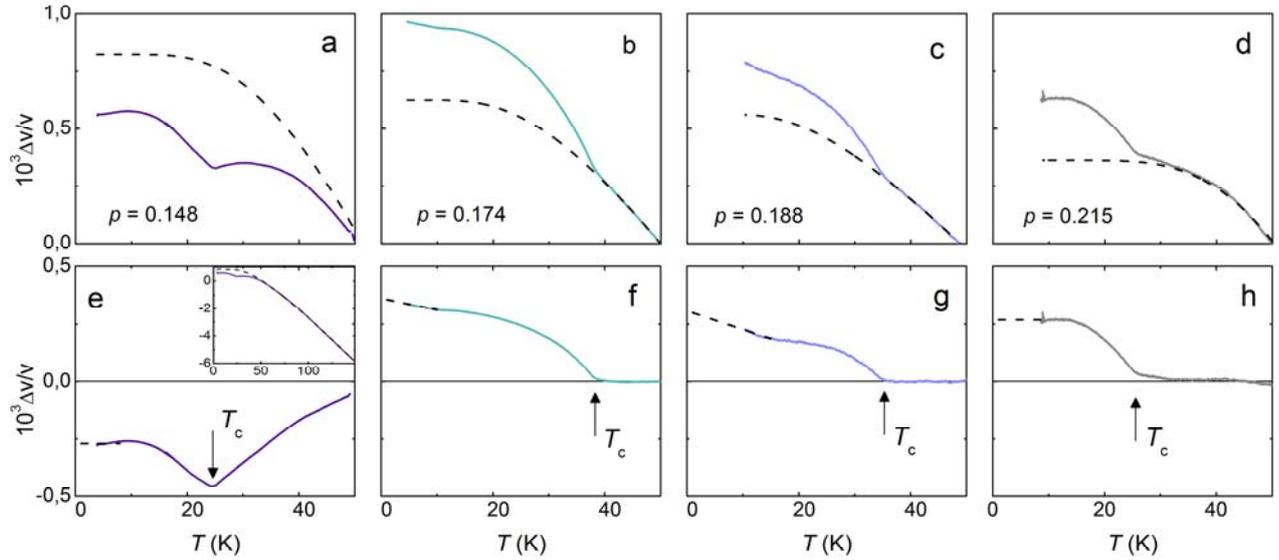

**Fig. S3. Temperature dependence of the zero field sound velocity.**

**a – d** – Raw temperature dependence of the sound velocity in zero magnetic field, $\Delta v/v(B=0)$, at different doping levels. Dashed lines are fits to the data using Eq. E2. This fit is used as a lattice background, $\Delta v/v(B=0)_{background}$, and it is subtracted to the data in order to extract the zero field electronic contribution to the sound velocity $\Delta v/v(B=0)_e$ (see Methods for details). **e – h** – Electronic contribution to the sound velocity $\Delta v/v(B=0)_e$. At all doping levels, a clear anomaly is observed at the superconducting $T_c$, as pointed with arrows. In the cuprates, the superconducting order parameter causes a hardening of the lattice as observed earlier (40). For $p = 0.148$ a softening is also visible above $T_c$. It is related to the slowing down of spin fluctuations, already present in zero magnetic field at this doping level. Because of this softening, Eq. E2 is fitted to the data of sample with $p = 0.148$ only for $T > 50$ K or so, as shown in the inset of panel **e**, while for other doping levels, it can be fitted for $T > T_c$. Dashed lines are linear extrapolation of the $\Delta v/v(B=0)_e$ used in Fig. 2 and 3. An upturn is observed upon cooling for $p = 0.174$ and $p = 0.188$, as previously reported in this doping range (40). For US sample with $p = 0.168$, a high quality zero field temperature dependence could not be obtained. We used data from a sample with similar doping $p = 0.174$ in order to estimate $\Delta v/v(B=0)_e$ for $p = 0.168$.

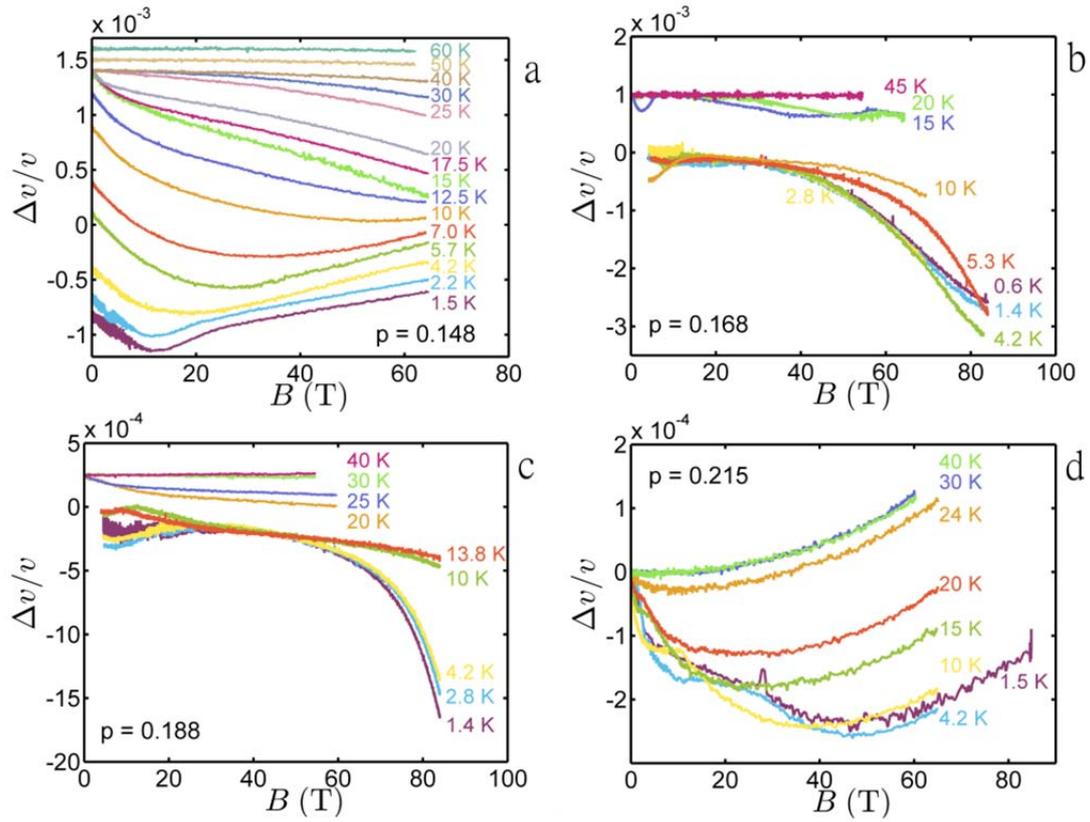

**Fig. S4. Pulsed field sound velocity data at all doping levels.**

**a – d** Field dependence of the sound velocity, $\Delta v/v(B)$, at different temperatures for $p = 0.148$ (panel a), $p = 0.168$ (panel b), $p = 0.188$ (panel c) and $p = 0.215$ (panel d). Curves are shifted vertically for clarity. All data presented here are from the downsweep part of the magnetic field pulse. From those data, cuts at constant magnetic field are made in order to obtain the temperature dependence of the field induced sound velocity.

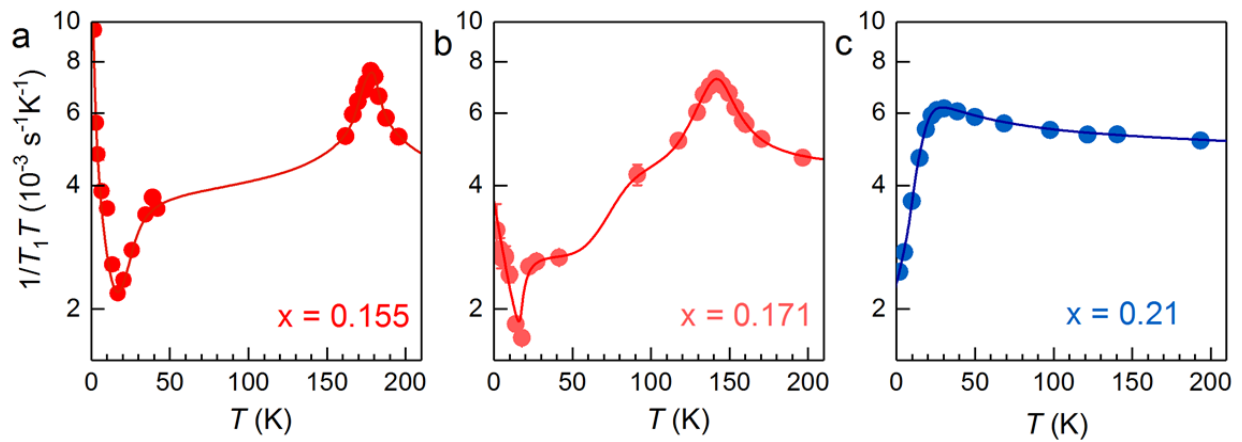

**Fig. S5. Contrasting NMR $T_1$ data below and above $p^* = 0.19$.**
a-c, Temperature dependence of $^{139}$La $1/T_1T$ in a field of 15 T for different doping levels. The peak in $1/T_1T$ around 180 K (x = 0.155) and 140 K (x = 0.171) is due to the tetragonal-to-orthorhombic structural transition (electric-field gradient fluctuations contributing to the nuclear relaxation through quadrupolar interaction), the drop below 40 K (all samples) is due to superconductivity and the low $T$ upturn is due to glassy slowing down. This latter is not observed for x = 0.21. Also, for this x = 0.21 sample, the $T$ dependence in the normal state as well as the large residual $1/T_1T$ value in the $T = 0$ limit are both due to the structural transition at $T \approx 6$ K (that is, the relaxation peak produced by electric-field gradient fluctuations becomes very broad). Lines are guides to the eye.

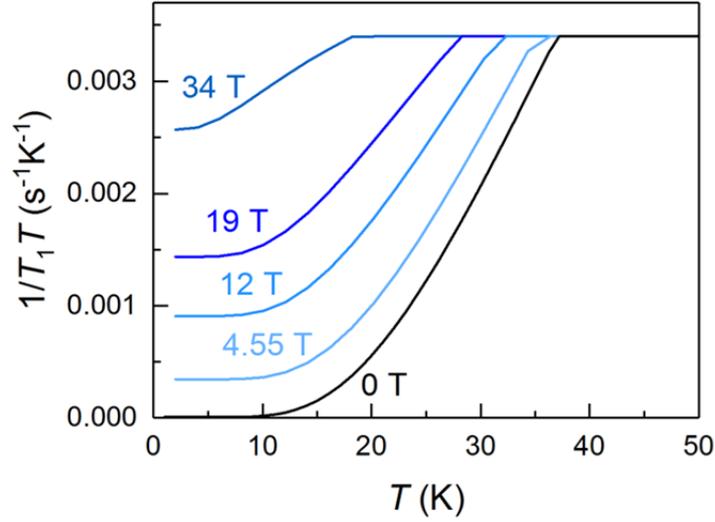

**Fig. S6. Model of field dependence of NMR $T_1$ due to superconductivity.**

The exact $B$ and $T$ dependence of the superconducting gap $\Delta$ is uncertain but, given the smallness of this correction, any function that mimics a gap which is gradually filled with increasing field and temperature is sufficient. For simplicity, we used an exponential decay of $T_1$ in the superconducting state, even though a power-law decrease is expected for a $d$-wave gap. We used the following simple function: $1/T_1T_{backgr.}(B,T) = c_0 + (c_1 - c_0)\exp\left(\frac{\Delta}{T_c}\right)\exp\left(-\frac{\Delta}{T}\right)$, where $c_0$ is the minimal relaxation rate and $c_1$ is the constant relaxation above $T_c$ (justified by the fact that experimental $1/T_1T$ is $T$ independent just above $T_c$). Both $\Delta$ and $T_c$ may depend on $B$ in a complicated way but we take the simplest possible expressions: $T_c(B) = T_{c,0}\sqrt{1 - \frac{B}{\mu_0 H_{c2}}}$ (mean-field approximation), $\Delta(B) = 2.15\, T_c(B)$ (effective $d$-wave gap, ref. 47), $c_0(B) = \frac{B}{\mu_0 H_{c2}}$ (linearly increasing minimal $1/T_1T$, ref. 48). Then, all that is needed to plot $1/T_1T_{backgr.}$ is the zero-field $T_c$ and $H_{c2}$. We use $H_{c2}$ values shown in Fig. S10. Above $T_c(B)$ the background is constant: $1/T_1T\,(T > T_c) = 0.0034$ s$^{-1}$K$^{-1}$.

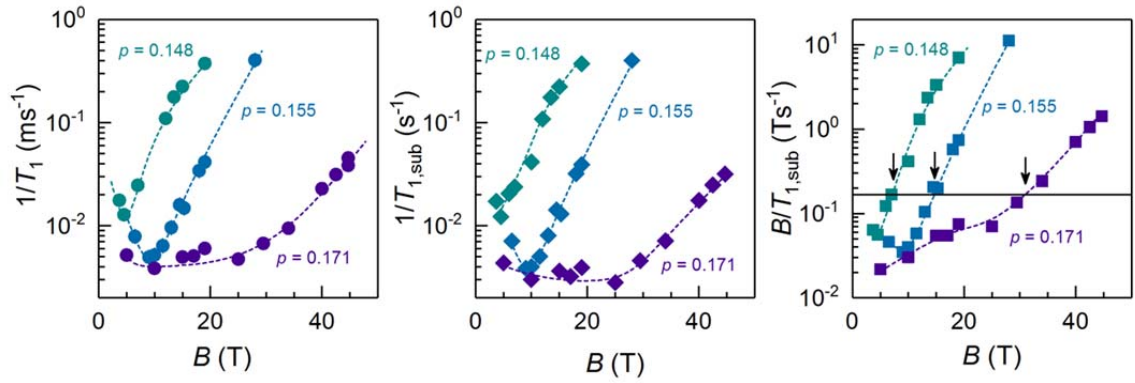

**Fig. S7. Determination of the field scale $B_{slow}$ in NMR data.**
**a,** Field dependence of $^{139}$La $1/T_1$ at $T = 1.7$ K for different doping levels. The minimum at low fields arises from the balance between an increase of $1/T_1$ upon increasing $B$ (field-induced spin freezing) and an unavoidable frequency effect in NMR ($1/T_1$ decreases with increasing NMR frequency, itself proportional to $B$). **b,** Relaxation rate $1/T_{1,\,sub}$ after subtraction of a background accounting for the field dependence of the superconducting gap (an example of which is shown in Fig. S6). **c,** $B/T_{1,sub}$ values. The multiplication by $B$ accounts for the unavoidable $1/B$ dependence of $1/T_1$ (see Methods). The $B_{slow}$ value for each sample (marked by arrows) is determined from the criterion $B/T_{1,\,sub} = 0.166$ T s$^{-1}$, chosen so as to match neutron scattering results, namely $B_{slow} = 7$ T for $x = 0.148$ (see Methods). Dashed lines are guides to the eye.

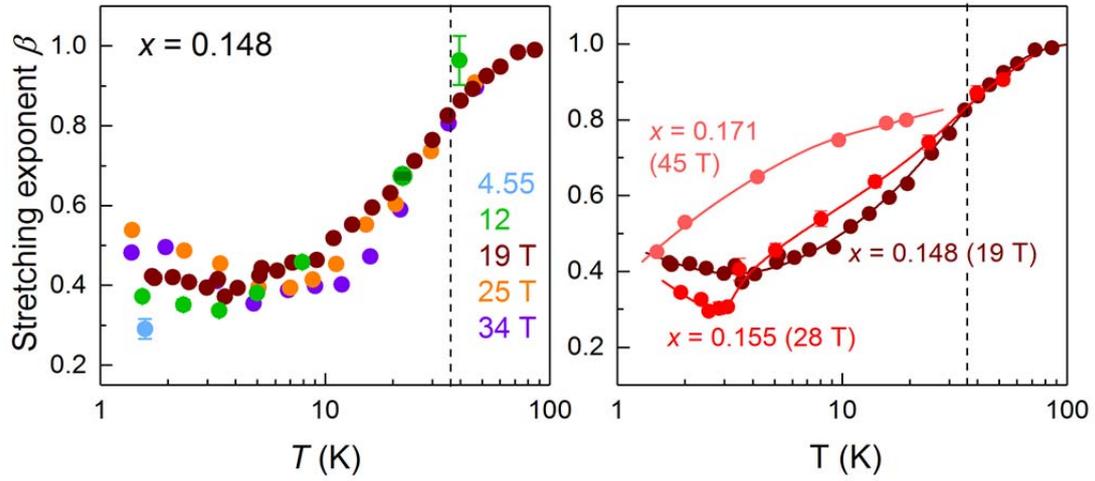

**Fig. S8. Stretching exponent $\beta$ in NMR $T_1$ measurements.**
**a,** stretching exponent $\beta$ for different fields in $La_{1.852}Sr_{0.148}CuO_4$. **b,** stretching exponent $\beta$ for Sr concentrations $x = p$. The stretching exponent provides a phenomenological measure of the width of the distribution of $T_1$ values (ref. 14 and references therein).

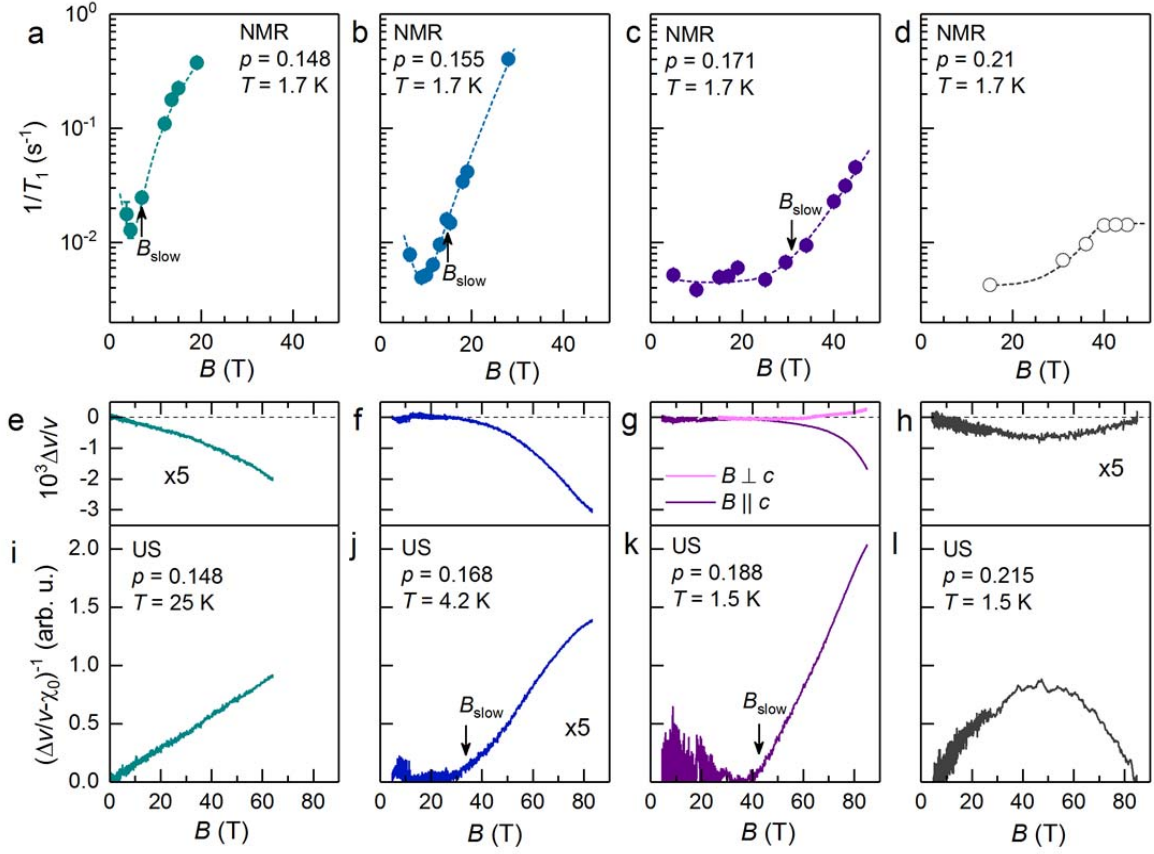

**Fig. S9. Field dependence of glassy freezing.**

**a-d**, Field dependence of $^{139}$La $1/T_1$ at $T = 1.7$ K for different doping levels. The minimum at low fields arises from the balance between an increase of $1/T_1$ upon increasing $B$ (field-induced spin freezing) and an unavoidable frequency effect in NMR ($1/T_1$ decreases with increasing NMR frequency, itself proportional to $B$). Dashed lines are guides to the eye. **e-h**, Field dependence of the sound velocity for different doping levels. In contrast to NMR, ultrasound is measured at constant frequency as a function of field. For each sample with a doping level $p < p^*$, the field dependence is plotted at a temperature where $\Delta v/v$ decreases upon cooling, *i.e.* for $T \geq T_{\min}$ (see Fig. 1). For $p = 0.215$, $\Delta v/v$ is plotted at the lowest $T$ achieved during the experiment. For $p < p^*$ and field $B \parallel c$, $\Delta v/v$ is almost field independent at low fields and above a doping-dependent onset field $B_{\text{slow}}$, it shows a strong $1/B$ dependence, highlighted in panels **i-l**. For $B \parallel (110)$ (panel **g**), $\Delta v/v$ shows no softening, and only increases up to 84 T. This highlights that the field effect arises from competition with superconductivity. **i-l**, Inverse field dependent sound velocity $(\Delta v/v - \chi_0)^{-1}$, where $\chi_0$ is a doping-dependent constant. For $p < p^*$, $(\Delta v/v - \chi_0)^{-1}$ is linear as a function of $B$ for $B > B_{\text{slow}}$ (pointed by arrows), with $B_{\text{slow}}$ increasing with doping, as shown in Fig. 4. At doping level $p = 0.168$ and for $T = 4.2$ K, $(\Delta v/v - \chi_0)^{-1}$ deviates from linearity for $B > 70$ T, which probably signals the proximity to spin freezing. For $p \approx 0.21$, both the weak field induced softening (**h,i**) and the field dependence of $1/T_1$ (**d**) stop at about 40 T, a field value consistent with the upper critical field $B_{c2}$ (Fig. S10). The field dependence below 40 T is explained by the suppression of superconductivity in both cases.

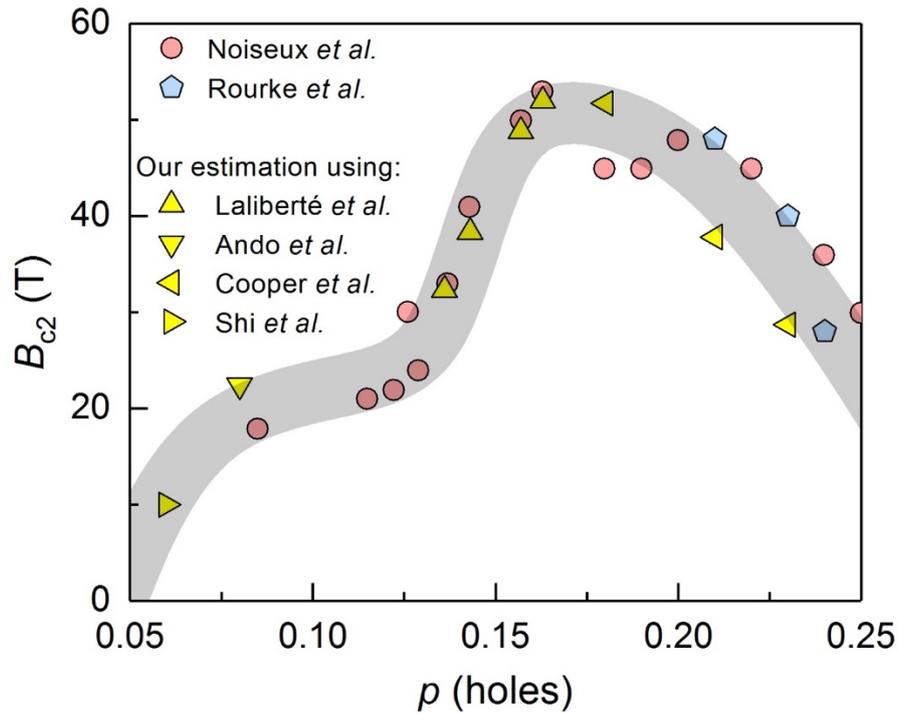

**Fig. S10. Estimation of $B_{c2}$ from transport measurements.**
The upper critical field $B_{c2}$ can be estimated by interpolation of the vortex melting line down to $T = 0$, as it has been done in YBCO (45) using the so-called Blatter formula (46). Since the formula is not exact close to $T = 0$ and the absence of a vortex liquid at such temperature is still an open question in $La_{2-x}Sr_xCuO_4$, these values should be seen only as a rough estimation. This method is used by Noiseux *et al.* (circles, ref. 44) and ourselves using transport data from the literature (triangles, refs. 25,27,41,42). The fields at which low temperature magnetoresistance deviates from $B^2$ dependence (43) is also reported.

| Technique | $T_c$ (K) | $T_{st}$ (K) | Doping $p$ (hole/ Cu) |
|---|---|---|---|
| NMR | 36.2 ± 1 | 194 ± 2 | 0.148 |
| NMR | 38.1 ± 1 | 177.5 ± 2 | 0.155 |
| NMR | 37.5 ± 1 | 141.5 ± 3 | 0.171 |
| NMR | 25.6 ± 1 | 6 ± 10 | 0.210 |
| US | 24.5 ± 1 | 194.5 ± 6 | 0.148 |
| US | 37.7 ± 1 | 149.3 ± 6 | 0.168 |
| US | 35.4 ± 1 | 104.3 ± 5 | 0.188 |
| US | 26.0 ± 1 | < 8 K | 0.215 |

**Table S1. $T_c$, $T_{st}$ and hole doping of NMR and US samples.**

For all US samples $T_c$ was detected as a change of slope in the temperature dependence of the mode $(c_{11}-c_{12})/2$ (see Fig. S3), except for $p = 0.168$ where it was better resolved in the longitudinal mode $c_{11}$. US sample with doping $p = 0.148$ features an anomalously low $T_c$ for such doping. However, this fact does not seem to affect the magnetic properties of the sample which appear to be characteristic of doping $p \approx 0.15$. Indeed, the temperature and field scales for the spin freezing of this US sample are very similar to the $p = 0.148$ NMR sample, which features $T_c = 36$ K.